# Artificial intelligence and democracy:

## Towards digital authoritarianism or a democratic upgrade?[*]

Fereniki Panagopoulou

Associate Professor at Panteion University

Ph.D. (Humboldt), M.P.H. (Harvard), LL.M., Ph.D. (N.K.U.A.).

## I) Introduction

Do robots vote? Do machines make decisions instead of us? No, (at least not yet), but this is something that could happen. The impact of Artificial Intelligence (AI)[1] on democracy is a complex issue that requires thorough research and careful regulation. At the most important level, that of the electoral process, it is noted that it is not determined by the AI, but it is greatly impacted by its multiple applications. New types of online campaigns, driven by AI applications, are replacing traditional ones. The potential for manipulating voters and indirectly influencing the electoral outcome should not be underestimated. Certainly, instances of voter manipulation are not absent from traditional political campaigns, with the only difference being that digital manipulation is often carried out without our knowledge, e.g. by monitoring our behavior on social media.[2] Nevertheless, we should not overlook the positive impact that AI has in the upgrading of democratic institutions by providing a forum for participation in decision-making.

In this context, as a first step, we look into the potential jeopardization of democratic processes posed by the use of AI tools. Secondly, we consider the possibility of strengthening democratic processes by using AI, as well as the democratization of AI itself through the possibilities it offers. And thirdly, the impact of AI on the representative system is also discussed. The paper is concluded with recommendations and conclusions.

## II) Risks posed for democracy

Misuse of AI tools can lead to the undermining of democratic political processes or the manipulation of individuals through specific targeting, which will destabilize democracy.[3] The potential risks are many and far from negligible.

---

[*] Warm thanks are expressed to Professor Georgios Giannopoulos, Professor of the School of Law of the University of Athens, for the very constructive dialogue with him.

[1] According to Article 3(3)(a). 1 of Regulation 2024/1689 "AI system" is a mechanical system designed to operate with different levels of autonomy and capable of exhibiting adaptability after implementation and which, for explicit or implicit purposes, infers from the input data it receives how to produce output data, such as predictions, content, recommendations or decisions that can affect physical or virtual environments.

[2] Fereniki Panagopoulou-Koutnatzi, *Artificial Intelligence: The Road Towards Digital Constitutionalism. An Ethico-Constitutional Consideration* (Athens: Papazisis Publications, 2023), 221 ff.

[3] Raluca Csernatoni, *Can Democracy Survive the Disruptive Power of AI?*, December 18, 2024, accessed 28 February 2025, https://carnegieendowment.org/research/2024/12/can-democracy-survive-the-disruptive-power-of-ai?lang=en

## 1. Misinformation

The operation of algorithms and the associations of persons generated by them are of considerable importance in the implementation of the democratic principle, in the sense of its majoritarian dimension. It is possible to produce purposeful (mis)information, a misrepresentation of reality,[4] into which one can easily slip without even realizing it, whilst being extremely difficult to get out of it, from a technical point of view. This kind of targeted information can affect the electoral outcome and, to a large extent, the democratic principle,[5] in the sense that the outcome of elections is not ultimately decided by the people, but by interest groups that have the technical capacity to influence the will of the people. In this sense, misinformation is a threat to liberal democracy and its institutions.[6] The provision of impartial information is crucial for the proper functioning of democracy and the undistorted expression of the will of the people when it comes to elections.[7] Electoral processes can be undermined through activities such as the dissemination of both deliberate and unintentional disinformation, the use of deep fake technology,[8] or the manipulation of individuals through specific targeting. The misuse of these technologies poses a significant threat to democracies, as it enables malicious actors — from political opponents to foreign adversaries — to manipulate public perceptions, disrupt electoral processes, and amplify disinformation.[9] Texts created by generative AI [10] are particularly dangerous in this regard. As these systems produce highly persuasive texts, they enable states and non-state actors to propagate disinformation and malicious narratives.[11]

Generative AI models played a major role in the 2024 US presidential election campaign, with fake images and deep fakes created by AI flooding social media platforms.[12] Fake images were published by both sides: Trump reposted a fake image created by AI, in which singer Taylor Swift was depicted endorsing his campaign — which is something she never did. Democrats also published fake AI pictures of Trump being arrested.[13]

## 2. Data exploitation-utilization

---

[4] See the relevant problematic in Lillian Mitrou, "Digital Democracy, Participation and Threats," in *Rule of Law and Democracy in the Digital Age*, ed. George Karavokiris (Athens: Hellenic Parliament Foundation for Parliamentarism and Democracy, 2024), 53 ff. (78 ff.).

[5] Council of Europe, *Study on the Human Rights Dimensions of Automated Data Processing Techniques (in Particular Algorithms) and Possible Regulatory Implications*, March 2018, accessed 28 February 2025, https://ec.europa.eu/futurium/en/european-ai-alliance/study-human-rights-dimensions-automated-data-processing-techniques-particular.html

[6] See Mitrou, "Digital Democracy," 78.

[7] See Papadopoulou, "False News and Intolerant Discourse," in *Rule of Law and Democracy in the Digital Age*, ed. Giorgos Karavokyris (Athens: Hellenic Parliament Foundation for Parliamentarism and Democracy, 2024), 99 ff. (117).

[8] In accordance with recital 134 and Article 3(3)(a) of Regulation (EC) No . 60 of the AI Regulation, "objectification products" are "image, sound or video content produced or manipulated by an AI which bears a resemblance to real persons, objects, places, entities or events and which may give the deceptive impression that it is genuine or real".

[9] See Papadopoulou, "False News and Intolerant Discourse," 117.

[10] Generative AI models learn the patterns and structure of their input training data and subsequently generate new data that have similar characteristics.

[11] See Papadopoulou, "False News and Intolerant Discourse," 117.

[12] Ibid.

[13] Ibid.

Data is very valuable, and this is something that justifies their characterization as "new gold".[14] Those who process it can better understand their constituents, optimize their actions, and make data-driven decisions.[15] It has always been necessary for governance to rely on data. Big data contributes to the effectiveness of public services and strategic planning, and influences interactions between citizens and public services, policies and administrative systems.[16] AI can collect and analyze this data in real time, training the algorithm [17] and enabling campaign strategists to adjust their approaches based on public opinion. Data collection can serve either as a tool for identifying the electorate's core needs or as a means of influencing and manipulating voters. By analyzing voters' unique psychographic and behavioral profiles on social media, AI can be used to sway individuals towards a particular candidate or spread hostility against opponents to influence voting decisions.[18] The creation of psychographic profiles and targeted messages, facilitated by Big Data, in online digital campaigns driven by deception and intimidation can influence a wide range of activities, from propaganda to policymaking.[19] This is because election campaigns are increasingly shifting online, and targeted ads on digital media are shaping election outcomes.[20]

Governance based on available data is neither neutral nor uncontroversial, as the data itself is neither fully accessible nor free from controversy. Data often contains biases.[21] The vast amounts of available data exceed human capacity for analysis, comprehension and, ultimately, practical use. As a result of the above, there is growing reliance on automated algorithms to identify patterns and support decision-making, further deepening our dependence on such technologies and exacerbating power imbalances.[22] Biases are not only shaped by data but also by algorithm design and AI training practices,[23] which can either amplify or reduce them.[24] Data inequality arises primarily from unequal access to data.[25] Even when databases are publicly available, only a

---

[14] See Tom Dausy, "Data, the New Gold: How AI Is Unlocking Insights and Driving Business Growth," *Medium*, May 19, 2024, accessed 28 February 2025, https://medium.com/@tomdausy/data-the-new-gold-how-ai-is-unlocking-insights-and-driving-business-growth-c458b5676f08

[15] Ibid.

[16] UNESCO, *Artificial Intelligence and Democracy* (2024), 13, accessed 28 February 2025, https://unesdoc.unesco.org/ark:/48223/pf0000389736

[17] See *Plan for Greece's Transition to the AI Era*, accessed 28 February 2025, https://foresight.gov.gr/wp-content/uploads/2024/11/Sxedio_gia_tin_metavasi_TN_Gr.pdf

[18] UNESCO, *Artificial Intelligence and Democracy*, 12.

[19] Ibid.

[20] See Cornelius Erfort, "Targeting Voters Online: How Parties' Campaigns Differ," *Electoral Studies* 92 (December 2024), accessed 28 February 2025, https://www.sciencedirect.com/science/article/pii/S0261379424001306?via%3Dihub

[21] A classic case of prejudice is the systematic discrimination against women who applied for technical jobs at Amazon, such as software engineering positions. The algorithm developed this kind of bias because Amazon's existing pool of software engineers is overwhelmingly male, and the new software was fed data on the resumes of these engineers. See Rachel Goodman, "Why Amazon's Automated Hiring Tool Discriminated Against Women," *ACLU*, October 12, 2018, accessed 28 February 2025, https://www.aclu.org/news/womens-rights/why-amazons-automated-hiring-tool-discriminated-against

[22] UNESCO, *Artificial Intelligence and Democracy*, 14.

[23] Cf. the COMPAS algorithm, Correctional Offender Management Profiling for Alternative Sanctions.

[24] UNESCO, *Artificial Intelligence and Democracy*, 14.

[25] The European Data Act (the Regulation on harmonized rules on fair access to and use of data - also known as the Data Act - entered into force on 11 January 2024) is a key pillar of the European data strategy that will make a significant contribution to achieving the Digital Decade goal of promoting digital transformation.

select few have the skills or resources or time to analyze, interpret, manage or utilize them.[26] The current Big Data ecosystem is causing significant inequalities, although this is a different form of poverty and wealth[27] - one not based on material goods but on information.[28] Essentially, there are three categories of people when it comes to databases: those who generate data, those with the capabilities and resources to store it, and those who know how to leverage its value. The latter form the largest and most privileged group, setting the rules that govern the use of and participation in Big Data.[29]

**3. Manipulation**

The Cambridge Analytica scandal demonstrated the potential misuse of AI in influencing electoral behavior. This is because internet users' data can be used to create a political, ideological, or psychological profile of a voter, allowing for the targeted delivery of personalized advertisements or automated messages to promote or discredit a particular candidate.

Firstly, there is the issue of undermining citizen autonomy: by creating a detailed psychographic profile and exploiting weaknesses, fears, and obsessions through personalized messages designed to trigger emotions, the citizen is subjected to a form of automated 'brainwashing', aimed at influencing behavior.[30] Secondly, internet users are rarely informed or asked for their consent regarding the use of their data for these purposes.[31] Thirdly, that raises concerns about the violation of the right to vote and the principle of equal opportunities for candidates to be elected.[32] Fourthly, algorithms are used to rapidly generate and spread false or distorted news as part of propaganda strategies designed to manipulate the information and emotions of internet users in a specific direction. For example, false or distorted news -including statements by political figures, during the election period, about alleged violent crimes committed by migrants

---

It stipulates that connected products must be designed and manufactured and related services must be provided in such a way that the data generated by these products and services are directly accessible to users (Article 3). If data cannot be made directly accessible, it must be made available on request without undue delay (Article 4).

There is an exception to the obligation to provide immediate access or availability of data upon request where this would undermine the security of the connected product, resulting in serious adverse effects on the health, safety or security of natural persons.

According to Articles 3 and 4, the data that must be readily accessible or available on request include product data, which is data generated by the use of the connected product and which is designed to be retrievable through an electronic communications service, physical connection or access to the device, service-related data, which is data representing the digitization of a user's actions (including actions within) and events related to the use of the connected product and which is designed to be retrievable through an electronic communications service, physical connection or access to the device, service-related data, which is data representing the digitization of a user's actions (including actions within) and events related to the use of the connected product, and data representing the digitization of a user's actions (including actions within) and events related to the use of the connected product.

[26] UNESCO, *Artificial Intelligence and Democracy*, 14.
[27] See "How Data Analytics Drive Growth for Wealth Management Firms," August 2, 2024, accessed 28 February 2025, https://www.uneecops.com/blog/data-analytics-for-wealth-management-firm/
[28] UNESCO, *Artificial Intelligence and Democracy*, 14.
[29] UNESCO, *Artificial Intelligence and Democracy*, 14.
[30] See Panagopoulou-Koutnatzi, *Artificial Intelligence: The Road to a Digital Constitutionalism*, 286.
[31] Ibid., 288.
[32] See Latanya McSweeney, *Psychographics, Predictive Analytics, Artificial Intelligence & Bots: Is the FTC Keeping Pace?*, accessed 28 February 2025, http://dataprivacylab.org/projects/onlineads/, 514 ff.

and asylum seekers -can be exploited in campaigns to fuel intense fear, intolerance, and xenophobia, ultimately promoting extreme ideologies.[33] Fifthly, the use of algorithms to generate deep fakes (videos in which people are digitally altered or replaced) in order to mislead the public and discredit political opponents poses an equally dangerous threat to democratic processes.

Furthermore, we must not overlook the fact that the delivery of concrete, premeditated, and tested knowledge risks undermining the liberal nature of our constitution, in the sense of imposing the 'average' of existing knowledge on science and thought more broadly, leading to a standardized perception of reality.[34] The use of AI systems to influence voters politically and shape the electoral outcome has raised serious concerns at both the European and international levels. The question arises as to whether legislative intervention is necessary to safeguard the core of liberal democracy. This issue is complex, as the entrenchment of a particular perception can, in turn, lead to the reinforcement of a particular electoral preference and, in this case, to the indirect manipulation of the electorate.

The risks associated with the use of AI in electoral contests can be numerous, difficult to address and, in some cases, intolerable (within the meaning of Article 5 of the AI Regulation (EU) 2024/1689), particularly when they result in the manipulation of citizens. It is suggested that legislators prioritize this issue by requiring digital platforms, developers, and algorithm distributors to ensure algorithmic transparency and by criminalizing the creation and dissemination of harmful deep fake content that leads to political manipulation.[35] At the same time, software developers and distributors must prevent the creation of harmful deep fake audio and video content and may be held liable if their preventive measures are easily circumvented.

This requires strengthening the mechanisms for verifying the reliability of news information (fact-checking, through the marking of content as true, false or controversial, or by using a special watermark).[36] Additionally, it calls for the development of specialized skills to combat the falsification of news, the improvement of media literacy and education, and the empowerment of stakeholders (citizens, journalists, and both private and public bodies) through the adoption of clear, responsible, fair and widely accepted rules of conduct and operation in the new, advanced technological environment.[37] The relevant debate is complex. A key question that arises is who exercises control and on what principles.[38] In this context, the Digital Service Act imposes transparency and accountability obligations regarding content moderation.[39]

**4. Towards a "privatization" of elections?**

---

[33] See Panagopoulou-Koutnatzi, *Artificial Intelligence: The Road to a Digital Constitutionalism*, 286.
[34] Fereniki Panagopoulou-Koutnatzi, "Legal and Ethical Considerations for the Use of ChatGPT in Education," *DITE 2023*, 6 ff.
[35] See *Plan for Greece's Transition to the AI Era*, 94.
[36] See Mitrou, "Digital Democracy," 83. But there is also the question of who will qualify the news as true, false or disputed.
[37] *Plan for Greece's Transition to the AI Era*, 94.
[38] See Mitrou, "Digital Democracy," 87.
[39] In this sense, this Act is the most important reform regarding digital platforms. See Ioannis Igglezakis, *The Law of the Digital Economy*, 2nd ed. (Athens-Thessaloniki: Sakkoulas Publications), 71.

The role of large private companies that control the Internet is crucial. Their infrastructure and algorithms determine how people communicate, what information is presented to participants,[40] and in what order. In short, private entities increasingly influence the regulation of social behavior and the flow of information.[41]

Almost every stage of AI model development — from computing infrastructure to training data — is controlled by an oligopoly of technology companies. Public oversight of the development of these systems and governance remains minimal. The concentration of AI development in monopolistic forces poses significant risks.[42] There is a need to explore alternative structures of AI governance and utilization that better serve the public interest. One such structure could be participatory schemes:[43] these schemes are employee-owned and operated enterprises with rules that correspond to the rich global experience of community-driven business practices that distribute both control and capital. Today, participatory schemes are estimated to employ nearly 10% of the world's population.[44] Participatory structures are designed to provide more equitable ownership and governance compared to investor-owned companies.[45] This structure is expected to mitigate the growing concerns of various stakeholders: consumers gain greater influence and a financial stake in the technological systems that shape their lives; businesses build trust with consumers and regulators while remaining aligned with public needs; and regulators facilitate a competitive marketplace with the potential to enhance the social impact of AI.[46]

**5. Algorithmic governance?**

To a large extent, decision-making tends to be algorithmic,[47] as automated systems now play a significant role in government decisions.[48] Addressing complex issues mandates algorithmic decision-making. Managing the complexity of modern societies without such processes is difficult to imagine, as they process vast amounts of information and automate tasks that would otherwise be impossible or less efficient.[49] Nevertheless, the key issue is the extent to which and how the use of automated decision-making systems is compatible with political decision-making.[50] AI systems can improve our understanding of social preferences and facilitate more objective evaluations of public policies. They are also useful in situations where there is a large amount of data and choices are categorized into binary classifications.[51] Even so, they are limited in cases of data scarcity or ambiguous situations where policy decisions are imperative and carry more certainty than any calculation. In any case, in a democracy, the final decision rests with the people who hold sovereignty, regardless of the extent of data processing.[52] Therefore, algorithms should assist, rather than replace, those who make political decisions.[53]

---

[40] See Nicolas Suzor, "Digital Constitutionalism: Using the Rule of Law to Evaluate the Legitimacy of Governance by Platforms," *Social Media + Society* (2018), accessed 28 February 2025, https://journals.sagepub.com/doi/10.1177/2056305118787812

[41] Ibid., 2.

[42] See Anton Korinek and Jai Vipra, "AI Monopolies," *Economic Policy*, March 27, 2024, accessed 28 February 2025, https://www.economic-policy.org/79th-economic-policy-panel-ai-monopolies/

[43] See Sarah Hubbard, *Cooperative Paradigms for Artificial Intelligence* (Cambridge, MA: Ash Center for Democratic Governance and Innovation, Harvard Kennedy School, November 20, 2024), accessed 28 February 2025, https://ash.harvard.edu/resources/cooperative-paradigms-for-artificial-intelligence/

[44] International Cooperative Alliance, *Co-ops Employ 10% of the Global Employed Population*, September 25, 2017, accessed 28 February 2025, https://ica.coop/en/media/news/co-ops-employ-10-global-employed-population

[45] See Connor Spelliscy, Sarah Hubbard, Nathan Schneider, and Samuel Vance-Law, "Toward Equitable Ownership and Governance in the Digital Public Sphere," *Stanford Journal of Blockchain Law & Policy*, 2024, accessed 28 February 2025, https://stanford-jblp.pubpub.org/pub/equitable-ownership-and-governance/release/1

[46] See Hubbard, *Cooperative Paradigms for Artificial Intelligence*.

[47] See Fereniki Panagopoulou, "Algorithmic Decision Making in Public Administration," in *Rule of Law and Democracy in the Digital Age*, ed. Giorgos Karavokyris (Athens: Foundation of the Hellenic Parliament for Parliamentarism and Democracy, 2024), 141 ff.

[48] See John Danaher, "The Threat of Algocracy: Reality, Resistance and Accommodation," *Philosophy & Technology* 29 (2016): 245–268.

[49] UNESCO, *Artificial Intelligence and Democracy*, 46.

**6. Have the risks been confirmed?**

In 2024, an unprecedented number of national elections took place worldwide.[54] It was the largest electoral event in human history, with 3.7 billion voters across 72 countries heading to the polls.[55] Despite early public concerns about AI having a dramatic impact on this year's elections, experts agreed that most of those fears did not materialize. Instead, a more nuanced reality is emerging—AI now permeates all aspects of elections, from campaign operations to the information ecosystem.[56] Many platforms reportedly took proactive measures to address election-related concerns, redirected political queries to authoritative sources in order to develop their own AI-based countermeasures to identify and address coordinated inauthentic behavior.[57]

In addition to the above, alarming practices involving bots have also been observed. "Bots" are social media accounts that are not owned by real individuals but are controlled by their operators for a specific purpose. In Romania, the European Commission launched a formal investigation into TikTok due to "serious indications" of foreign interference in the country's recent presidential election. The second-round vote was cancelled after intelligence documents revealed that 25,000 TikTok bot accounts were suddenly activated just weeks before the first-round polls opened. The European Union's executive has opened a formal investigation into TikTok due to "serious indications" of foreign interference in Romania's recent presidential election using the video-sharing platform. The accounts supported independent candidate Georgescu, who described Russia's Vladimir Putin as a "patriot and leader", although he denied being one of his supporters.[58] The EU regulatory authorities will assess whether TikTok's advertising policies and the systems it uses to recommend content to users violate the Digital Services Act (DSA), which aims to prevent the spread of misinformation and curb illegal activities on the Internet.[59]

---

[50] Ibid.

[51] Ibid.

[52] Ibid.

[53] See Panagopoulou, "Algorithmic Decision Making in Public Administration," 180.

[54] Adrien Morissard, "It's the Biggest Election Year in Modern History. Will Democracy Prevail?" *NPR*, July 3, 2024, accessed 28 February 2025, https://www.npr.org/2024/07/03/1198912778/its-the-biggest-election-year-in-modern-history-will-democracy-prevail

[55] See Bruce Schneier and Nathan Sanders, *The Apocalypse That Wasn't: AI Was Everywhere in 2024's Elections, but Deepfakes and Misinformation Were Only Part of the Picture* (Cambridge, MA: Ash Center for Democratic Governance and Innovation, Harvard Kennedy School, December 4, 2024), accessed 28 February 2025, https://ash.harvard.edu/articles/the-apocalypse-that-wasnt-ai-was-everywhere-in-2024s-elections-but-deepfakes-and-misinformation-were-only-part-of-the-picture/

[56] See Sarah Hubbard, *The Role of AI in the 2024 Elections* (Cambridge, MA: Ash Center for Democratic Governance and Innovation, Harvard Kennedy School, December 5, 2024), accessed 28 February 2025, https://ash.harvard.edu/resources/the-role-of-ai-in-the-2024-elections/

[57] Ibid.

[58] See Alex Loftus, "EU Investigates TikTok over Alleged Russian Meddling in Romanian Vote," *BBC News*, December 17, 2024, accessed 28 February 2025, https://www.bbc.com/news/articles/cm2v13nz202o

This matter is regulated, firstly, by the Digital Service Act, which aims to prevent the spread of misinformation and to curb illegal activities online. The Act in question has as its main objective to create a safer online environment for digital users and digital companies, as well as to protect fundamental rights in the digital domain by establishing new rules to combat illegal content on the internet, including products, services and information, in full compliance with the Charter of Fundamental Rights. Secondly, if artificial intelligence is used for their dissemination, bots are considered a prohibited risk, falling also under the EU AI Act. Thirdly, if personal data is processed, then the General Data Protection Regulation is also applicable.

**III) Upgrading democratic institutions**

The question arises as to whether facilitating access to AI-enabled digital governance applications has a positive effect on democracy.[60] There is increasing discussion about the "platformization of democracy", as a significant portion of debates take place on platforms that, while facilitating them, simultaneously shape them in various ways.[61]

**1. Facilitating participation in consultation processes**

AI tools can strengthen democracy by facilitating citizen participation in consultation processes. The innovative use of AI has the potential to enhance deliberative democracy at multiple levels of decision-making, from national legislation to local government and corporate governance. This is because AI facilitates the consultation process by summarizing consultations and aggregating vast amounts of incoming online information.

AI can help scale up consultation and democratic participation. Among the many tasks that AI tools can perform are selecting and allocating participants in various kinds of assemblies (the LEXIMIN algorithm[62] is already available and in use), facilitating consultations (Meta's community forums pioneered this function), summarizing consultations, and aggregating vast amounts of online input (the pol.is system[63] is currently being used by the Taiwanese government to this end).[64]

In this way, AI can also serve as a key tool for promoting the democratic governance of AI technology itself. Indeed, ensuring that AI is developed for the benefit of society, while upholding transparency and accountability, requires the participation of society itself, which deliberative democracy can help facilitate.[65]

The above are also linked to building "digital trust" for platforms, which can be achieved through continuous public evaluation, as well as democratic representation and participation where critical decisions are made.[66] Developing AI for the benefit of society, while ensuring transparency and accountability, cannot take place without the involvement of society itself or, at the very least, credible representatives of the people.

---

[59] Ibid.
[60] See the relevant problematic in Mitrou, "Digital Democracy," 55 ff.
[61] See Tarleton Gillespie, "The Politics of 'Platforms,'" *New Media & Society* 12, no. 3 (2010): 347–364.
[62] See Sylvain Bouveret and Michel Lemaître, "Computing Leximin-Optimal Solutions in Constraint Networks," *Artificial Intelligence* 173 (2009): 343–364, https://doi.org/10.1016/j.artint.2008.10.010
[63] Colin Megill, "pol.is in Taiwan," *Pol.is Blog*, accessed 28 February 2025, https://blog.pol.is/pol-is-in-taiwan-da7570d372b5
[64] See *Plan for Greece's Transition to the AI Era*, 94.
[65] Ibid.

A possible experiment in democratic representation could involve the creation of citizens' assemblies.[67] These assemblies have become an established mechanism through which various countries, including Ireland,[68] France,[69] Belgium[70] and Canada,[71] seek to address legitimacy deficits and governance challenges. These are large bodies of citizens, usually convened either in person or virtually, to deliberate on complex policy issues with the aim of producing policy recommendations and, in some cases, even legislative proposals. A recent OECD report documented hundreds of such assemblies (or similar models) worldwide.[72] In essence, a citizens' assembly consists of representatives drawn from adult citizens, residents of a given region, or even individuals residing outside the territory.[73] These representatives are selected through a quasi-random process but are deliberately 'stratified' based on characteristics such as nationality, ethnic origin, gender, and social class, and deliberate on an issue.[74] The assembly follows three structured stages of debate: It begins with an educational phase, where its members are introduced to the issue at hand. This is followed by a stage in which assembly members listen to diverse perspectives from stakeholders, individuals, or collectives. It concludes with the deliberation, during which the assembly votes on proposals.[75]

---

[66] Ibid.

[67] In the sense of citizenship as a broader concept than nationality, which is important for documenting the views of groups of people who do not have the nationality of the state in which they are located.

[68] Citizens' assemblies have been set up in Ireland to assess possible amendments to the Constitution (e.g. on same-sex marriage and reproductive rights).

[69] This is the Citizens' Climate Convention in France, which met in 2019/2020. The Convention originates from the "Gilets Jaunes" (Yellow Vests) revolt that started in late 2018 in response to the introduction of a new fuel tax by the French government. As the French government struggled to respond to the revolt (which was supported across the political spectrum), it sought to channel popular energy - and perhaps contain it - by organizing a 'Grand National Debate'. The National Debate took the form of a series of local Citizens' Assembly-type assemblies. Out of this experience came the idea of holding a national climate policy authority. In July 2019, the government entrusted the Assembly with the task of making recommendations on how France could reduce carbon emissions in a socially equitable way by 40% below 1990 levels by 2030.

[70] In Belgium, a permanent body with the characteristics of a citizens' assembly has been proposed which would have the power to decide whether to put a proposal to a referendum. Organisation for Economic Co-operation and Development (OECD), *Eight Ways to Institutionalise Deliberative Democracy* (Paris: OECD Publishing, 2021), accessed 28 February 2025, https://www.oecd.org/gov/open-government/eight-ways-to-institutionalise-deliberative-democracy.htm

[71] In Canada, a citizens' assembly was created to reform electoral legislation. See https://nationalcitizensassembly.ca/

[72] Organisation for Economic Co-operation and Development (OECD), *Innovative Citizen Participation and New Democratic Institutions: Catching the Deliberative Wave* (Paris: OECD Publishing, 2020), accessed 28 February 2025, https://www.oecd-ilibrary.org/fr/governance/innovative-citizen-participation-and-new-democratic-institutions_339306da-en

[73] See Stuart White, *Citizens' Assemblies and Democratic Renewal* (Athens: ENA Institute for Alternative Policies, April 2022), 3, accessed March 3, 2025, https://enainstitute.org/publication/synelefseis-politon-dimokratiki-ana/

[74] Ibid.

[75] Ibid.

To maximize legitimacy, these citizens' assemblies should engage with the broader public through various online and offline consultation mechanisms, some of which could be uniquely facilitated by "open" AI tools. It should be noted that openness is essential not only for the sustainable and ethical development of AI technologies but also for ensuring conditions of digital sovereignty.

Selecting people from all walks of life through AI tools seems to have three main advantages. Firstly, assemblies achieve greater demographic representation compared to elected representation. This is because elected representatives tend to come disproportionately from the most privileged groups in society—overwhelmingly male, white and middle class, for example. This imbalance leads to political inequality and a lack of diverse perspectives that emerge when people from a broader social spectrum come together.[76] Secondly, assemblies are seen as a means of bridging the trust gap between politicians and citizens, provided that citizens trust their representatives in the assembly and that politicians genuinely consider its proposals.[77] Thirdly, citizens' assemblies can contribute to enhancing the quality of direct democracy.[78] It should not be overlooked, however, that assemblies also carry the risk of 'majoritarian tyranny', as they may be susceptible to the disproportionate influence of wealthy individuals who can more easily finance petition and referendum campaigns.[79]

At this point it is crucial to emphasize that citizen participation in the decision-making process is not limited to organized structures, like citizens' assemblies. It can also occur in a more dispersed and spontaneous way, allowing individuals to participate in online consultation platforms on their own initiative. While spontaneous participation lowers barriers to engagement, it also carries the risk of mass involvement by individuals who have no reason to participate in decision-making, as the decisions do not concern them.

**2. Facilitating political campaigns**

**a. Translation tools**

One of the beneficial uses of AI is the linguistic translation of election campaigns.[80] Local governments in Japan[81] and California,[82] as well as prominent politicians, such as Indian Prime

---

[76] Ibid., 4.
[77] Ibid., 5.
[78] Ibid., 6.
[79] Ibid.
[80] See Ananya Bhattacharya, "Political Campaigns Embrace AI to Reach Voters Across Language Barriers: Advocacy Groups Like AAPI Victory Alliance Are Using ChatGPT and Claude to Reach Speakers of Hindi, Tagalog, and More," *Rest of World*, September 19, 2024, https://restofworld.org/2024/aapi-victory-alliance-ai-voter-outreach/
[81] *Japan Today*, "Japanese Mayor Suddenly Speaks Fluent English with AI Video That Surprises Even Him," June 7, 2024, accessed 28 February 2025, https://japantoday.com/category/politics/japanese-mayor-suddenly-speaks-fluent-english-with-ai-video-that-surprises-even-him
[82] See Zhe Wu, "As Bay Area Cities Adopt Real-Time AI Translation for Public Meetings, SF Abstains: Community Groups Say San Francisco Needs to Improve Language Access at City Hall," *San Francisco Public Press*, December 6, 2024, accessed 28 February 2025, https://www.sfpublicpress.org/as-bay-area-cities-adopt-real-time-ai-translation-for-public-meetings-sf-abstains/

Minister Narenda Modi[83] and New York City Mayor Eric Adams,[84] have used AI to translate meetings and speeches for their diverse constituencies.[85]

Even when politicians themselves do not speak through AI, their voters may rely on it to listen to them. Google launched free translation services for an additional 110 languages this summer, making real-time translations accessible to billions of people via their smartphones.[86]

**b. Virtual chats**

Other candidates leveraged AI-powered chat features to connect with voters. In the U.S. presidential primaries, some politicians used AI chatbots[87] of themselves as part of their campaigns.[88] Fringe candidate Jason Palmer defeated Joe Biden in the American Samoa primary, at least in part due to his use of AI-generated emails, texts, audio, and video. Pakistan's former Prime Minister Imran Khan used an AI-generated clone of his voice to deliver speeches from prison.[89]

Perhaps the most effective use of this technology occurred in Japan, where an independent candidate for Tokyo governor, Takahiro Anno, used an AI avatar[90] to answer 8,600 voter questions, securing fifth place in a highly competitive field of 56 candidates.[91]

Chatbots also encourage users to leave comments and share valuable insights about their opinions.[92]

**c. Political organizing**

On the political organizing side, AI assistants are used for a variety of purposes, such as helping to craft political messages and strategy, creating advertisements, drafting speeches, and coordinating efforts to gather and elect voters.[93] In 2023 in Argentina, both incumbent presidential candidates used it to develop posters, videos and other campaign materials.[94]

---

[83] *The Times of India*, "BJP to Use AI to Translate PM's Speeches," March 8, 2024, accessed 28 February 2025, https://timesofindia.indiatimes.com/city/bengaluru/bjp-to-use-ai-to-translate-pms-speeches/articleshow/108318912.cms

[84] See Anthony Izaguirre, "Adams' Revelation That He Uses AI to Speak in Mandarin Stirs Outcry: 'The Mayor Is Making Deep Fakes of Himself,'" *Fortune*, October 17, 2023, accessed 28 February 2025, https://fortune.com/2023/10/17/new-york-city-mnayor-eric-adams-uses-ai-to-speak-mandarin/

[85] See Schneier and Sanders, *The Apocalypse That Wasn't*.

[86] Isaac Caswell, "110 New Languages Are Coming to Google Translate," *Google Blog*, accessed 28 February 2025, https://blog.google/products/translate/google-translate-new-languages-2024/

[87] A chatbot is a software tool or web-based interface designed to engage in text or voice-based conversations.

[88] See Schneier and Sanders, *The Apocalypse That Wasn't*.

[89] Yaqiu Zhuang, "Imran Khan's 'Victory Speech' from Jail Shows A.I.'s Peril and Promise," *The New York Times*, February 11, 2024, accessed 28 February 2025, https://www.nytimes.com/2024/02/11/world/asia/imran-khan-artificial-intelligence-pakistan.html

[90] An avatar is a virtual character that someone creates to represent themselves online—in chats, games, and other digital spaces. It can take the form of a digital image, a pseudonym, or even a fully developed virtual persona.

[91] See Gideon Lichfield, "Meet Your AI Politician of the Future: A Young, No-Name Outsider Is Quietly Revolutionizing Japanese Political Campaigning," *Future Polis*, October 4, 2024, accessed 28 February 2025, https://futurepolis.substack.com/p/meet-your-ai-politician-of-the-future

[92] UNESCO, *Artificial Intelligence and Democracy*, 46.

In 2024, similar capabilities were utilized in various elections around the world. In the US, for example, a Georgia politician[95] used them to produce blog posts, campaign images, and podcasts.[96] Other AI systems help in advising candidates who seek to run for higher offices.[97]

Furthermore, AI can collect and analyze this data in real time, enabling campaign strategists to change their approaches based on public opinion.[98]

**IV) Transformation of representative democracy**

**1. The challenge**

Citizen participation in consultation processes, whether organized in the form of assemblies or occurring in a more dispersed manner, raises questions on the changing or evolving nature of representative democracy. New Internet technologies that expand opportunities for participation with the assistance of AI applications amplify the broader challenges involved. Therefore, it is essential to examine whether these new capabilities provided by the internet alter the qualitative characteristics of democratic participation in a way that impacts the functioning of the constitution.[99]

AI tools allow politicians to easily, quickly and effectively gauge public opinion on a wide range of issues within their mandate. This naturally raises the question of when a politician should go back and consult the public, on what types of issues, and through what means.[100] Modern online media have fostered new prospects and hopes for enhancing citizen participation in the decision-making process.[101] Nevertheless, this prompts the broader question: how is representative democracy being transformed by new internet technologies? While these technologies expand the pool of participants in consultation processes, they also impose new limitations.[102]

**2. The potential expansion of the consultation cycle**

---

[93] See Schneier and Sanders, *The Apocalypse That Wasn't*.
[94] Jack Nicas and Lucía Cholakian Herrera, "Argentina the First A.I. Election?" *The New York Times*, November 15, 2023, accessed 28 February 2025, https://www.nytimes.com/2023/11/15/world/americas/argentina-election-ai-milei-massa.html
[95] See Ross Williams, "Georgia Political Campaigns Start to Deploy AI but Humans Still Needed to Press the Flesh," *GPB News*, April 25, 2024, accessed 28 February 2025, https://www.gpb.org/news/2024/04/25/georgia-political-campaigns-start-deploy-ai-humans-still-needed-press-the-flesh
[96] Podcasts are digital audio files that are made available online.
[97] See Sean J. Miller, "AI Is Helping Candidates Decide on Runs for Higher Office," *Campaigns & Elections*, October 22, 2024, accessed 28 February 2025, https://campaignsandelections.com/campaigntech/ai-is-helping-candidates-decide-on-runs-for-higher-office/
[98] UNESCO, *Artificial Intelligence and Democracy*, 12.
[99] Vasiliki Christou, "Towards a Digital Municipality," in *Rule of Law and Democracy in the Digital Age*, ed. Giorgos Karavokyris, 12 -25 (Athens: Hellenic Parliament Foundation for Parliamentarism and Democracy, 2024), 19 ff., esp. 24.
[100] Ibid., p. 19 ff.
[101] Ibid., 23.
[102] Ibid., 29.

The expansion lies in the fact that participation in the consultation process is made very easy through AI applications and, as a result, it is accessible to a wider range of people, who can participate in consultations from home. Moreover, as a rule, it takes place without any kind of identification of participants. In theory, everyone can participate in any consultation, whether it concerns them or not. This widens the circle of people who decide on any given issue. At the same time, however, this can also lead to the involvement of citizens who have no reason to participate.

This potential concern pertaining to unconditional participation in consultation mechanisms is prevalent in many modern consultation media, where strict participant identification measures are in place. Nevertheless, such identification raises concerns about the secrecy of the procedure and the protection of the participants' personal data. At the same time, unrestricted access carries the risk of involving groups that have no reason to be making decisions. This naturally leads to the question: on which matters are citizens legitimately entitled to decide? Are all Greeks abroad, for example, all Greeks abroad legitimately entitled to in the decision-making process on Greece's internal affairs?

Could we take this idea a step further by allowing Greeks to participate in decision-making on German or even American affairs? Any expansion of the consultation cycle could be justified on the grounds that important decisions made by one country have repercussions on others. For example, immigration restrictions in Germany affect Greece, which may be forced to absorb more immigrants as a result. The withdrawal of the US from the Paris Agreement, along with its broader climate and environmental policies, has consequences for other countries. Similarly, the abolition of English-language university programs in the Netherlands will push students to seek studies in other countries. Notwithstanding the above, the expansion of those participating in the consultation cycle could also signal an attempt to shape or influence final outcomes. This risk arises when a large group of citizens is deliberately steered to make a particular choice.

In any case, it can be concluded that electronic media have the potential to expand the cycle of consultation, whether in a strict form (with an identification system) or in a loose form (without one).

**3. Risk of restriction/exclusion of participants**

Participation in decision-making through AI tools may restrict participants, depriving the digitally illiterate of the ability to participate in processes they cannot comprehend. This is also happening with the digitization of the state (gov.gr), the banking sector (e-banking) and the provision of health services (e-health). There is talk of a kind of digital authoritarianism, in the dual sense that: F*irstly*, the internet not only compels participation in online processes; and, *secondly*, it also dictates the terms of engagement.[103] Certain segments of the population remain unrepresented on these platforms due to various systemic inequalities (such as gender, age, socio-economic status, or lack of connectivity itself, leading to disconnection and complete digital exclusion).[104] This creates inequality between the online underprivileged and the online affluent, reinforcing a digital disparity that mirrors the broader distinction between the haves and have-nots.[105] Indeed, since secure consultation processes require strong identification, the necessary digital literacy extends well beyond the novice level. Nevertheless, this gap is expected to narrow over time.

---

[103] See Csernatoni, *Can Democracy Survive the Disruptive Power of AI?*
[104] UNESCO, *Artificial Intelligence and Democracy*, 13.
[105] See Panagopoulou, *Electronic Voting*, 126.

### 4. Is representative democracy ultimately changing?

The above challenges lead us to the question whether expanding the consultation cycle truly addresses the shortcomings of representative democracy, namely, its detachment from the people and the threat of technocracy.[106] Does it not, however, create the illusion of citizen participation in the public sphere, when in reality their involvement has no effect?[107] And does this not shift political responsibility onto the people who are, by definition, not in a position to bear it? Who will bear the responsibility if a misguided choice leads the country to economic disaster? Will the people face political consequences? At this point, it is crucial to emphasize that AI should serve as a tool for consultation and not as a means for diffusing political responsibility. This distinction is essential to preserving the structural foundations of representative democracy. Representatives will listen to the people – but the final decision will remain in their hands.

### V) Abstention or conditional acceptance?

There are two possible choices: abstention or conditional acceptance. The first one is technophobic and utopian, meaning that it is non-existent as an option: the genie is already out of the bottle.[108] The second one requires careful policymaking. In this context, policymakers should consider the trade-offs of AI content identification, which is a technique that embeds a unique, traceable signature in AI-generated content, marking it as an AI product. Visible watermarks promote transparency but may interfere with artistic intent, whereas digital watermarks, embedded in metadata, are more subtle but easier to manipulate.[109] The global, interconnected nature of online content suggests that broader, harmonized standards across jurisdictions may be necessary for effective multilateral governance. In this context, companies have been urged to develop credible mechanisms, such as watermarking.[110] To this end, the AI Regulation imposes obligations on AI providers and developers to ensure transparency, as well as the detection and tracking of AI-generated material (Recital 115). Addressing the challenges posed by AI-generated content will require coordination among a wide range of stakeholders, including governments, AI companies, social media platforms, and users. Technology companies play a key role in developing authentication and provenance tools to identify and trace the origin of AI-generated content.[111] While using AI to combat AI offers some promise, many detection tools remain inaccessible to the public.[112]

Detection technology alone may fall short without regulatory oversight and efforts to improve public digital literacy. Enhancing AI and information literacy is essential, making digital skills education an urgent priority.[113]

### V. Recommendations

---

[106] Christou, "Towards a Digital Municipality," 37.
[107] Christou, "Towards a Digital Municipality," 41.
[108] Schneier and Sanders, *The Apocalypse That Wasn't*.
[109] See Csernatoni, *Can Democracy Survive the Disruptive Power of AI?*
[110] European Commission, *Hiroshima Process International Guiding Principles for Advanced AI System*, 30 October 2023, accessed 28 February 2025, https://digital-strategy.ec.europa.eu/en/library/hiroshima-process-international-guiding-principles-advanced-ai-system
[111] Ibid.
[112] Ibid.
[113] Ibid.

Ensuring the peaceful integration of AI while enhancing democracy requires a series of measures.

*Firstly*, there is an urgent need for education and raising awareness regarding the challenges of AI.

*Secondly*, careful regulation is required to ensure that AI is used for the "common good", guided by humanitarian principles such as diversity, equality and inclusion, which are embedded in the protection of human rights, democracy, and the rule of law.[114] The legal framework must establish independent oversight mechanisms for effective compliance to ensure accountability. Such an oversight mechanism can only be effective if it is proactive and engaged from the outset, rather than reacting after issues arise.[115] Indeed, while introducing sanctions for non-compliant behavior is important, relying solely on ex-post sanctions and fines — which large private companies can often afford to pay, regardless of the amount — may not achieve the desired results. This is because, once a particular AI technology has been introduced and used, reversing its effects or "undoing the damage" is often difficult – if not impossible — regardless of whether it complies with ethical standards, human rights, democracy, and the rule of law.[116] Additionally, efforts should focus on democratizing[117] data.[118] Alongside legal measures to prevent or restrict the spread of harmful rhetoric, fact-checking institutions have proven effective in mitigating its negative impact.[119] To ensure maximum transparency, which is essential for informed public opinion, states are encouraged to promote codes of best practice for companies and mandate the recognition of AI-generated content as measures to combat misinformation.

*Thirdly*, promoting transparency and explainability in AI systems is essential to help people understand decision-making processes and the criteria behind their outcomes.[120] Issues of representation, exclusion, and discrimination in all politics in general may be further exacerbated when decisions are made by AI systems that those affected cannot understand. Transparency should also be tied to the definition of anonymity in social media,[121] ensuring that readers are not misled as to the identity of the author.[122]

*Fourthly*, the principle of pluralism[123] should be upheld throughout the AI development process. This includes ensuring gender diversity among professionals, inclusive system design, curated datasets, mitigation of biased exclusion policies, fair facial recognition procedures, and balanced information recommendations should be ensured. Pluralism should also guide the democratization of AI governance by involving new actors — such as regions, cities, private entities and citizens — in decision-making processes.[124]

---

[114] UNESCO, *Artificial Intelligence and Democracy*, 19.
[115] Ibid.
[116] UNESCO, *Artificial Intelligence and Democracy*, 20.
[117] See Hippolyte Lefebvre, Christine Legner, and Elizabeth A. Teracino, "5 Pillars for Democratizing Data at Your Organization," *Harvard Business Review*, November 24, 2024, accessed 28 February 2025, https://hbr.org/2023/11/5-pillars-for-democratizing-data-at-your-organization
[118] UNESCO, *Artificial Intelligence and Democracy*, 19.
[119] UNESCO, *Artificial Intelligence and Democracy*, 19.
[120] UNESCO, *Artificial Intelligence and Democracy*, 19.
[121] La Moncloa, "Pedro Sánchez Announces New Initiatives to Prevent 'the Digital Space from Becoming the Wild West,'" February 5, 2025, accessed 28 February 2025, https://www.lamoncloa.gob.es/lang/en/presidente/news/Paginas/2025/20250205-digital-rights-observatory.aspx
[122] See Fereniki Panagopoulou-Koutnatzi, *On Freedom: The New Technologies as a National, European, and International Challenge to the Freedom to Disseminate Ideas* (Athens-Thessaloniki: Sakkula Publications, 2010), 112.
[123] Catharina Rudschies, Ingrid Schneider, and Judith Simon, "Value Pluralism in the AI Ethics Debate - Different Actors, Different Priorities," *International Review of Information Ethics*, accessed 28 February 2025, https://informationethics.ca/index.php/irie/article/view/419/396

*Fifthly*, AI relies on the digitization of the state in order to function.[125] Digitization strategies should not only focus on technological transformation and economic modernization but also incorporate objective with direct democratic implications. These include the digitization of public administrations and the promotion of digital literacy[126] among citizens, all rooted in democratic values such as equality, inclusion, and accountability.[127] The democratic principles of equality, accountability, and transparency[128] should be the cornerstone of the implementation of AI systems, particularly when it comes to public services.[129]

*Sixthly*, there is an urgent need to establish universal standards for digitization and artificial intelligence.[130] These standards should incorporate diverse perspectives, interests, and objectives from stakeholders worldwide, with special consideration for marginalized regions (for example, Roma people).[131]

**VI) Conclusion**

AI tools can foster new forms of participation, whilst also facilitating the spread of misinformation. They have the potential to enhance the accountability of public institutions and their representatives, promote greater participation and pluralism to enrich citizen participation and make democracy more inclusive and adaptable. At the same time, they can also reinforce authoritarian tendencies and be used for potentially malicious and manipulative purposes.[132] The complexity of the issue calls for in-depth research to understand these threats and opportunities, enabling the development of policies that minimize threats and maximize opportunities. Supervisory authorities must effectively monitor technologies used in democratic participation, without resorting to technophobia. Further to the above, advanced disclosure methods are essential to ensure transparency in distinguishing between human and AI interactions.[133] Ultimately, it is our responsibility to shape AI as a tool for democratic upgrading, not digital authoritarianism.

---

[124] UNESCO, *Artificial Intelligence and Democracy*, 19.
[125] See Panagopoulou, "Algorithmic Decision Making in Public Administration," 179.
[126] See *Plan for Greece's Transition to the AI Era*, 84.
[127] UNESCO, *Artificial Intelligence and Democracy*, 21.
[128] See *Plan for Greece's Transition to the AI Era*, 14.
[129] UNESCO, *Artificial Intelligence and Democracy*, 21.
[130] Cf. Article 40 of the Rules of Procedure for AI.
[131] UNESCO, *Artificial Intelligence and Democracy*, 21.
[132] UNESCO, *Artificial Intelligence and Democracy*, 10.
[133] European Commission, *Hiroshima Process International Guiding Principles for Advanced AI Systems*, 12.

**References**

Bhattacharya, Ananya. "Political Campaigns Embrace AI to Reach Voters Across Language Barriers: Advocacy Groups Like AAPI Victory Alliance Are Using ChatGPT and Claude to Reach Speakers of Hindi, Tagalog, and More." *Rest of World*, September 19, 2024. https://restofworld.org/2024/aapi-victory-alliance-ai-voter-outreach/

Bouveret, Sylvain, and Michel Lemaître. "Computing Leximin-Optimal Solutions in Constraint Networks." *Artificial Intelligence* 173 (2009): 343–364. https://doi.org/10.1016/j.artint.2008.10.010

Caswell, Isaac. "110 New Languages Are Coming to Google Translate." *Google Blog*. Accessed 28 February 2025. https://blog.google/products/translate/google-translate-new-languages-2024/

Christou, Vasiliki. "Towards a Digital Municipality." In *Rule of Law and Democracy in the Digital Age*, edited by Giorgos Karavokyris, 12–25. Athens: Hellenic Parliament Foundation for Parliamentarism and Democracy, 2024.

Council of Europe. *Study on the Human Rights Dimensions of Automated Data Processing Techniques (in Particular Algorithms) and Possible Regulatory Implications.* March 2018. Accessed 28 February 2025. https://ec.europa.eu/futurium/en/european-ai-alliance/study-human-rights-dimensions-automated-data-processing-techniques-particular.html

Csernatoni, Raluca. *Can Democracy Survive the Disruptive Power of AI?* December 18, 2024. Accessed 28 February 2025. https://carnegieendowment.org/research/2024/12/can-democracy-survive-the-disruptive-power-of-ai?lang=en

Danaher, John. "The Threat of Algocracy: Reality, Resistance and Accommodation." *Philosophy & Technology* 29 (2016): 245–268.

Dausy, Tom. "Data, the New Gold: How AI Is Unlocking Insights and Driving Business Growth." *Medium*, May 19, 2024. Accessed 28 February 2025. https://medium.com/@tomdausy/data-the-new-gold-how-ai-is-unlocking-insights-and-driving-business-growth-c458b5676f08

Erfort, Cornelius. "Targeting Voters Online: How Parties' Campaigns Differ." *Electoral Studies* 92 (December 2024). Accessed 28 February 2025. https://www.sciencedirect.com/science/article/pii/S0261379424001306?via%3Dihub

European Commission. *Hiroshima Process International Guiding Principles for Advanced AI System.* 30 October 2023. Accessed 28 February 2025. https://digital-strategy.ec.europa.eu/en/library/hiroshima-process-international-guiding-principles-advanced-ai-system

Gillespie, Tarleton. "The Politics of 'Platforms.'" *New Media & Society* 12, no. 3 (2010): 347–364.

Goodman, Rachel. "Why Amazon's Automated Hiring Tool Discriminated Against Women." *ACLU*, October 12, 2018. Accessed 28 February 2025. https://www.aclu.org/news/womens-rights/why-amazons-automated-hiring-tool-discriminated-against

"How Data Analytics Drive Growth for Wealth Management Firms." August 2, 2024. Accessed 28 February 2025. https://www.uneecops.com/blog/data-analytics-for-wealth-management-firm/

Hubbard, Sarah. *Cooperative Paradigms for Artificial Intelligence.* Cambridge, MA: Ash Center for Democratic Governance and Innovation, Harvard Kennedy School, November 20, 2024. Accessed 28 February 2025. https://ash.harvard.edu/resources/cooperative-paradigms-for-artificial-intelligence/

Hubbard, Sarah. *The Role of AI in the 2024 Elections.* Cambridge, MA: Ash Center for Democratic Governance and Innovation, Harvard Kennedy School, December 5, 2024. Accessed 28 February 2025. https://ash.harvard.edu/resources/the-role-of-ai-in-the-2024-elections/

Igglezakis, Ioannis. *The Law of the Digital Economy.* 2nd ed. Athens-Thessaloniki: Sakkoulas Publications.

International Cooperative Alliance. *Co-ops Employ 10% of the Global Employed Population.* September 25, 2017. Accessed 28 February 2025. https://ica.coop/en/media/news/co-ops-employ-10-global-employed-population

Izaguirre, Anthony. "Adams' Revelation That He Uses AI to Speak in Mandarin Stirs Outcry: 'The Mayor Is Making Deep Fakes of Himself.'" *Fortune*, October 17, 2023. Accessed 28 February 2025. https://fortune.com/2023/10/17/new-york-city-mnayor-eric-adams-uses-ai-to-speak-mandarin/

*Japan Today*. "Japanese Mayor Suddenly Speaks Fluent English with AI Video That Surprises Even Him." June 7, 2024. Accessed 28 February 2025. https://japantoday.com/category/politics/japanese-mayor-suddenly-speaks-fluent-english-with-ai-video-that-surprises-even-him

Korinek, Anton, and Jai Vipra. "AI Monopolies." *Economic Policy*, March 27, 2024. Accessed 28 February 2025. https://www.economic-policy.org/79th-economic-policy-panel/ai-monopolies/

La Moncloa. "Pedro Sánchez Announces New Initiatives to Prevent 'the Digital Space from Becoming the Wild West.'" February 5, 2025. Accessed 28 February 2025. https://www.lamoncloa.gob.es/lang/en/presidente/news/Paginas/2025/20250205-digital-rights-observatory.aspx

Lefebvre, Hippolyte, Christine Legner, and Elizabeth A. Teracino. "5 Pillars for Democratizing Data at Your Organization." *Harvard Business Review*, November 24, 2024. Accessed 28 February 2025. https://hbr.org/2023/11/5-pillars-for-democratizing-data-at-your-organization

Gideon Lichfield, "Meet Your AI Politician of the Future: A Young, No-Name Outsider Is Quietly Revolutionizing Japanese Political Campaigning," *Future Polis*, October 4, 2024, accessed 28 February 2025, https://futurepolis.substack.com/p/meet-your-ai-politician-of-the-future.

Loftus, Alex. "EU Investigates TikTok over Alleged Russian Meddling in Romanian Vote." *BBC News*, December 17, 2024. Accessed 28 February 2025. https://www.bbc.com/news/articles/cm2v13nz202o

McSweeney, Latanya. *Psychographics, Predictive Analytics, Artificial Intelligence & Bots: Is the FTC Keeping Pace?* Accessed 28 February 2025. http://dataprivacylab.org/projects/onlineads/

Megill, Colin. "pol.is in Taiwan." *Pol.is Blog*. Accessed 28 February 2025. https://blog.pol.is/pol-is-in-taiwan-da7570d372b5

Miller, Sean J. "AI Is Helping Candidates Decide on Runs for Higher Office." *Campaigns & Elections*, October 22, 2024. Accessed 28 February 2025. https://campaignsandelections.com/campaigntech/ai-is-helping-candidates-decide-on-runs-for-higher-office/

Mitrou, Lilian. "Digital Democracy, Participation and Threats." In *Rule of Law and Democracy in the Digital Age*, edited by George Karavokiris, 53 ff. (78 ff.). Athens: Hellenic Parliament Foundation for Parliamentarism and Democracy, 2024.

Morissard, Adrien. "It's the Biggest Election Year in Modern History. Will Democracy Prevail?" *NPR*, July 3, 2024. Accessed 28 February 2025. https://www.npr.org/2024/07/03/1198912778/its-the-biggest-election-year-in-modern-history-will-democracy-prevail

Nicas, Jack, and Lucía Cholakian Herrera. "Argentina the First A.I. Election?" *The New York Times*, November 15, 2023. Accessed 28 February 2025. https://www.nytimes.com/2023/11/15/world/americas/argentina-election-ai-milei-massa.html

Organisation for Economic Co-operation and Development (OECD). *Eight Ways to Institutionalise Deliberative Democracy.* Paris: OECD Publishing, 2021. Accessed 28 February 2025. https://www.oecd.org/gov/open-government/eight-ways-to-institutionalise-deliberative-democracy.htm

Organisation for Economic Co-operation and Development (OECD), *Innovative Citizen Participation and New Democratic Institutions: Catching the Deliberative Wave* (Paris: OECD Publishing, 2020), accessed 28 February 2025, https://www.oecd-ilibrary.org/fr/governance/innovative-citizen-participation-and-new-democratic-institutions_339306da-en

Panagopoulou, Fereniki. "Algorithmic Decision Making in Public Administration." In *Rule of Law and Democracy in the Digital Age*, edited by Giorgos Karavokyris, 141 ff. Athens: Foundation of the Hellenic Parliament for Parliamentarism and Democracy, 2024.

Panagopoulou-Koutnatzi, Fereniki. *Artificial Intelligence: The Road to a Digital Constitutionalism: An Ethico-Constitutional Perspective.* Athens: Papazisis Publications, 2023.

Panagopoulou-Koutnatzi, Fereniki. "Legal and Ethical Considerations for the Use of ChatGPT in Education." *DITE* (2023): 6 ff.

Panagopoulou-Koutnatzi, Fereniki. *On Freedom: The New Technologies as a National, European, and International Challenge to the Freedom to Disseminate Ideas.* Athens-Thessaloniki: Sakkoulas Publications, 2010.

Papadopoulou, Lina. "False News and Intolerant Discourse." In *Rule of Law and Democracy in the Digital Age*, edited by Giorgos Karavokyris, 99 ff. (117). Athens: Hellenic Parliament Foundation for Parliamentarism and Democracy, 2024.

*Plan for Greece's Transition to the AI Era.* Accessed 28 February 2025. https://foresight.gov.gr/wp-content/uploads/2024/11/Sxedio_gia_tin_metavasi_TN_Gr.pdf

*Regulation (EU) 2023/2854 (The European Data Act).* Accessed 28 February 2025. https://eur-lex.europa.eu/eli/reg/2023/2854

Rudschies, Catharina, Ingrid Schneider, and Judith Simon. "Value Pluralism in the AI Ethics Debate - Different Actors, Different Priorities." *International Review of Information Ethics*. Accessed 28 February 2025. https://informationethics.ca/index.php/irie/article/view/419/396

Schneier, Bruce, and Nathan Sanders. *The Apocalypse That Wasn't: AI Was Everywhere in 2024's Elections, but Deepfakes and Misinformation Were Only Part of the Picture.* Cambridge, MA: Ash Center for Democratic Governance and Innovation, Harvard Kennedy School, December 4, 2024. Accessed 28 February 2025. https://ash.harvard.edu/articles/the-apocalypse-that-wasnt-ai-was-everywhere-in-2024s-elections-but-deepfakes-and-misinformation-were-only-part-of-the-picture/

Spelliscy, Connor, Sarah Hubbard, Nathan Schneider, and Samuel Vance-Law. "Toward Equitable Ownership and Governance in the Digital Public Sphere." *Stanford Journal of Blockchain Law & Policy*, 2024. Accessed 28 February 2025. https://stanford-jblp.pubpub.org/pub/equitable-ownership-and-governance/release/1

Suzor, Nicolas. "Digital Constitutionalism: Using the Rule of Law to Evaluate the Legitimacy of Governance by Platforms." *Social Media + Society* (2018). Accessed 28 February 2025. https://journals.sagepub.com/doi/10.1177/2056305118787812

*The Times of India*. "BJP to Use AI to Translate PM's Speeches." March 8, 2024. Accessed 28 February 2025. https://timesofindia.indiatimes.com/city/bengaluru/bjp-to-use-ai-to-translate-pms-speeches/articleshow/108318912.cms

UNESCO. *Artificial Intelligence and Democracy.* 2024. Accessed 28 February 2025. https://unesdoc.unesco.org/ark:/48223/pf0000389736

White, Stuart. *Citizens' Assemblies and Democratic Renewal*. Athens: ENA Institute for Alternative Policies, April 2022. Accessed March 3, 2025. https://enainstitute.org/publication/synelefseis-politon-dimokratiki-ana/

Williams, Ross. "Georgia Political Campaigns Start to Deploy AI but Humans Still Needed to Press the Flesh." *GPB News*, April 25, 2024. Accessed 28 February 2025. https://www.gpb.org/news/2024/04/25/georgia-political-campaigns-start-deploy-ai-humans-still-needed-press-the-flesh

Wu, Zhe. "As Bay Area Cities Adopt Real-Time AI Translation for Public Meetings, SF Abstains: Community Groups Say San Francisco Needs to Improve Language Access at City Hall." *San Francisco Public Press*, December 6, 2024. Accessed 28 February 2025. https://www.sfpublicpress.org/as-bay-area-cities-adopt-real-time-ai-translation-for-public-meetings-sf-abstains/

Zhuang, Yaqiu. "Imran Khan's 'Victory Speech' from Jail Shows A.I.'s Peril and Promise." *The New York Times*, February 11, 2024. Accessed 28 February 2025. https://www.nytimes.com/2024/02/11/world/asia/imran-khan-artificial-intelligence-pakistan.html

Zhuang, Yaqiu. "Imran Khan's 'Victory Speech' from Jail Shows A.I.'s Peril and Promise." *The New York Times*, February 11, 2024. Accessed 28 February 2025. https://www.nytimes.com/2024/02/11/world/asia/imran-khan-artificial-intelligence-pakistan.html